\documentclass[showpacs,preprintnumbers]{revtex4}
\usepackage{amsmath}
\usepackage{graphicx}

\begin{document}

\textbf{Comment on ``Classical and Quantum Interaction of the Dipole''}%
\bigskip

Tomislav Ivezi\'{c}

\textit{Ru%
\mbox
{\it{d}\hspace{-.15em}\rule[1.25ex]{.2em}{.04ex}\hspace{-.05em}}er Bo\v
{s}kovi\'{c} Institute, P.O.B. 180, 10002 Zagreb, Croatia}

\textit{ivezic@irb.hr\bigskip }

In [1], Anandan has presented a covariant treatment of the interaction of
the electric and magnetic dipole moments of a particle with the
electromagnetic field. Our aim is to make some important changes of the
results from [1]. Instead of dealing with component form of tensors $E^{\mu }
$, .. [1], we shall deal with tensors as four-dimensional (4D) geometric
quantities, $E^{a}$, .. . For simplicity, only the standard basis \{$e_{\mu
};\ 0,1,2,3$\} of orthonormal 4-vectors, with $e_{0}$ in the forward light
cone, will be used.

Anandan states: ``In any frame $D^{0i}$ and $D^{ij}$ that couple,
respectively, to the electric field components $F_{0i}$ and the magnetic
field components $F_{ij}$ are called the components of the electric and
magnetic dipole moments.'' Then, he defines that $d^{\mu }$ and $m^{\mu }$
are the components of $D^{\mu \nu }$, Eq. (2), and similarly that $E^{\mu }$
and $B^{\mu }$ are the components of $F^{\mu \nu }$, Eq. (4). Several
objections can be raised to such treatment.

It is proved in [2] that the primary quantity for the whole electromagnetism
is $F^{ab}$ (i.e., in [2], the bivector $F$). $F^{ab}$ can be decomposed as $%
F^{ab}=(1/c)(E^{a}v^{b}-E^{b}v^{a})+\varepsilon ^{abcd}v_{c}B_{d}$, whence $%
E^{a}=(1/c)F^{ab}v_{b}$ and $B^{a}=(1/2c^{2})\varepsilon ^{abcd}F_{bc}v_{d}$%
, with $E^{a}v_{a}=B^{a}v_{a}=0$; only three components of $E^{a}$ and $B^{a}
$ in any basis are independent. The 4-velocity $v^{a}$ is interpreted as the
velocity of a family of observers who measure $E^{a}$ and $B^{a}$ fields. $%
E^{a}$ and $B^{a}$ depend not only on $F^{ab}$ but on $v^{a}$ as well. In
the frame of ``fiducial'' observers, in which the observers who measure $%
E^{a}$, $B^{a}$ are at rest, $v^{a}=ce_{0}$. That frame with the $\{e_{\mu
}\}$ basis will be called the $e_{0}$-frame. In the $e_{0}$-frame $%
E^{0}=B^{0}=0$ and $E^{i}=F^{i0}$, $B^{i}=(1/2c)\varepsilon ^{ijk0}F_{jk}$.
In any other inertial frame, the ``fiducial'' observers are moving, and $%
v^{a}=ce_{0}=v^{\prime \mu }e_{\mu }^{\prime }$; under the passive Lorentz
transformations (LT), $v^{\mu }e_{\mu }$ transforms as any other 4-vector
transforms. The same holds for $E^{a}$ and $B^{a}$, e.g., $E^{a}=E^{\mu
}e_{\mu }=[(1/c)F^{i0}v_{0}]e_{i}=E^{\prime \mu }e_{\mu }^{\prime
}=[(1/c)F^{\prime \mu \nu }v_{\nu }^{\prime }]e_{\mu }^{\prime }$. $E^{\mu }$
transform by the LT again to the components $E^{\prime \mu }$ of the same
electric field. There is no mixing with the components of the magnetic
field. $E^{\prime \mu }$ are not determined only by $F^{\prime \mu \nu }$
but also by $v^{\prime \mu }$.

Only in the $e_{0}$-frame, and thus not in any frame, are $F^{i0}$ and $%
F^{jk}$ the electric and magnetic field components, respectively. The
assertion that, e.g., in any inertial frame it holds that $E^{0}=E^{\prime
0}=0$, $E^{i}=F^{i0}$, and $E^{\prime i}=F^{\prime i0}$, leads to the usual
transformations of the 3-vector $\mathbf{E}$, see, e.g., [3], Eq. (11.149).
In [4], the fundamental results are achieved that these usual
transformations of the 3-vectors $\mathbf{E}$ and $\mathbf{B}$ are not
relativistically correct and have to be replaced by the LT of the electric
and magnetic fields as 4D geometric quantities.

The electric and magnetic dipole moment 4-vectors $d^{a}$ and $m^{a}$,
respectively, can be determined from dipole moment tensor $D^{ab}$ in the
same way as $E^{a}$ and $B^{a}$ are obtained from $F^{ab}$; $%
D^{ab}=(1/c)(u^{a}d^{b}-u^{b}d^{a})+(1/c^{2})\varepsilon ^{abcd}u_{c}m_{d}$,
whence $d^{a}=(1/c)D^{ba}u_{b}$, and $m^{a}=(1/2)\varepsilon
^{abcd}D_{bc}u_{d}$, with $d^{a}u_{a}=m^{a}u_{a}=0$. $u^{a}=dx^{a}/ds$ is
the 4-velocity of the particle.

The whole discussion about $E^{a}$, $B^{a}$ and $F^{ab}$ can be repeated for
$d^{a}$, $m^{a}$ and $D^{ab}$. Now, only in the rest frame of the particle
and the $\{e_{\mu }\}$ basis, $u^{a}=ce_{0}$ and $d^{0}=m^{0}=0$, $%
d^{i}=D^{0i}$, $m^{i}=(c/2)\varepsilon ^{ijk0}D_{jk}$.

It is also stated in [1]: ``The electric and magnetic fields in the rest
frame ... .'' But, there is no rest frame for fields. The whole discussion
in [1] has to be changed using different 4-velocities $v^{a}$ and $u^{a}$.
Thus Eqs. (7) and (6) become $%
(1/2)F_{ab}D^{ba}=(1/c)D_{a}u^{a}+(1/c^{2})M_{a}u^{a}$, and $%
D_{a}=d^{b}F_{ba}$, $M_{a}=m^{b}F_{ba}^{\ast }$. Instead of Eq. (5), we have
that $(1/2)F_{ab}D^{ba}$ is the sum of two terms $%
(1/c^{2})[((E_{a}d^{a})+(B_{a}m^{a}))(v_{b}u^{b})-(E_{a}u^{a})(v_{b}d^{b})-(B_{a}u^{a})(v_{b}m^{b})]
$ and $(1/c^{3})[\varepsilon
^{abcd}(v_{a}E_{b}u_{c}m_{d}+c^{2}d_{a}u_{b}v_{c}B_{d})]$; the second term
contains the interaction of $E_{a}$ with $m^{a}$, and $B_{a}$ with $d^{a}$.
This last result significantly influences Eq. (17), and it will give new
interpretations for, e.g., the Aharonov-Casher and the R\"{o}ntgen phase
shifts. \bigskip

\noindent \textbf{References\medskip }

\noindent \lbrack 1] J. Anandan, Phys. Rev. Lett. \textbf{85}, 1354 (2000).

\noindent \lbrack 2] T. Ivezi\'{c}, Found. Phys. Lett. \textbf{18}, 401
(2005).

\noindent \lbrack 3] J.D. Jackson, \textit{Classical Electrodynamics}
(Wiley, New York, 1977) 2nd ed.

\noindent \lbrack 4] T. Ivezi\'{c}, Found. Phys. \textbf{33}, 1339 (2003)%
\textbf{; }Found. Phys. Lett. \textbf{18,} 301 (2005); Found.

Phys. \textbf{35,} 1585 (2005).

\end{document}